\def\be{\begin{equation}}
\def\ee{\end{equation}}
\def\bea{\begin{eqnarray}}
\def\eea{\end{eqnarray}}
\def\bse{\begin{subequations}}
\def\ese{\end{subequations}}

\documentclass[prl,twocolumn,showpacs,amsmath,amssymb]{revtex4}
\usepackage{bm}
\usepackage{latexsym}
\usepackage{dcolumn}
\usepackage[dvips]{graphicx}
\usepackage{amssymb}
\usepackage{graphics}
\usepackage{amsmath}
\usepackage{epsf}
\begin{document}
\title{Valley dependent many-body effects in 2D semiconductors}
\author{S. Das Sarma , E. H.\ Hwang, and Qi Li}
\affiliation{Condensed Matter Theory Center, Department of Physics,
  University of Maryland, College Park, Maryland 20742}
\date{\today}

\pacs{71.10.Ca,73.20.Mf}

\begin{abstract}
We calculate the valley degeneracy
($g_v$) dependence
of the many-body renormalization of quasiparticle properties in
multivalley 2D semiconductor structures due to the Coulomb
interaction between the carriers. Quite
unexpectedly, the $g_v$ dependence of many-body effects is
nontrivial and non-generic, and depends qualitatively on the specific
Fermi liquid property under consideration. While the interacting 2D
compressibility manifests monotonically increasing many-body
renormalization with increasing $g_v$, the 2D spin susceptibility
exhibits an interesting non-monotonic $g_v$ dependence with the
susceptibility increasing (decreasing) with $g_v$ for smaller (larger)
values of $g_v$ with the renormalization effect peaking around
$g_v\sim 1-2$. Our theoretical results provide a clear conceptual
understanding of recent valley-dependent 2D susceptibility
measurements in AlAs quantum wells.
\end{abstract}

\maketitle

Two dimensional (2D) electron (or hole) systems based on
semiconductors, e.g. inversion layers, FETs, quantum wells, and
heterostructures, have served as useful laboratory systems for studying
electron-electron interaction induced many-body effects for almost
forty years \cite{ando}. The reason is the ease with which the 2D carrier
density ($n$) can be tuned in these systems, thus effectively
changing the 2D electron gas (2DEG) from being a weakly interacting
system at large $n$ to a strongly interacting system at low $n$.
Such a variation in carrier density can not be  achieved in 3D metals
where only a modest tuning of the electron-electron interaction effect
is possible by going from one metal to another with the concomitant
complication of a
varying background lattice structure \cite{mahan}.
The electron interaction strength in a many-body Coulomb
quantum system is characterized \cite{mahan} by the dimensionless
$r_s$ parameter 
which measures the average inter-particle separation in units of the
effective Bohr radius. For the 2DEG, $r_s\equiv(\pi
n)^{-1/2}/({\kappa \hbar^2/me^2})$, where $n$,$m$, and $\kappa$ are
respectively the 2D electron density, the electron effective mass,
and the background lattice dielectric constant. Often $r_s$ is considered
to be the ratio of the average Coulomb potential energy, $e^2/(\kappa
r_0)$ where $r_0=(\pi n)^{-1/2}$ is the average inter-particle separation,
to the average kinetic energy $E_F=n\pi\hbar^2/m$ for a
2DEG (including the spin degeneracy). Although the appropriate
definition of $r_s$, also called the Wigner-Seitz radius, is in terms
of a dimensionless length, the definition in terms of the
dimensionless interaction energy is physically more appealing since
$r_s>1$ (or $r_s<1$) region may be considered as the dilute
strongly interacting (or the dense weakly interacting) regime. Varying
the carrier density one can increase $r_s$, thus accessing the strongly
interacting dilute regime where the quasi-particle renormalization
correction due to many-body effects would be large. In practice, the
very strongly correlated regime of $r_s\approx 10\sim 30$ \cite{lilly}
can be achieved
in 2DEG whereas in 3D metals\cite{mahan}, $r_s\approx 2\sim 5$. It
is, therefore, not surprising that 2D semiconductor systems have
long served as the laboratory test-bed for studying $r_s$ dependent
interaction effects. In general, the quasi-particle many-body
renormalization  is enhanced with increasing $r_s$, and in principle,
there could be quantum phase transitions (e.g Wigner crystallization,
ferromagnetic instability, dispersion instability, etc.) in the
strong-coupling $r_s\gg1$ regime.

Although electron interaction effects at $T=0$ are completely
characterized by the single dimensionless density parameter $r_s$ in
a single-valley system (e.g. GaAs), a multi-valley semiconductor
(e.g. Si, Ge, AlAs) is far more complex since the valley degeneracy
($g_v$), i.e the number of the equivalent valleys the electrons
occupy in the ground state due to the semiconductor band structure,
becomes an additional relevant parameter characterizing the
electron-electron interaction strength. (We consider only equal
valley population of all $g_v$ valleys and equal spin population
of spin up-down levels at $T=0$ in this work, i.e. we consider only
a paramagnetic ground state in both valley and spin quantum
numbers.) It is obvious that in the presence of a valley degeneracy
(i.e. an arbitrary value of $g_v>1$), both $r_s$ and $g_v$ will
determine the many-body renormalization effects, but a careful
investigation of how $g_v$ itself affects the quantum many-body
effects in 2DEG has not yet been carried out in the theoretical
literature \cite{ying}. This is precisely what we report in this Letter,
concentrating on the interacting 2D spin susceptibility and 2D
compressibility and calculating their many-body renormalizations as
functions of both $g_v$ and $r_s$.

A relatively straightforward interpretation of the $r_s$
parameter as a dimensionless coupling energy, i.e, the Coulomb
energy divided by the Fermi energy, gives $r_e\equiv {\rm Coulomb \;
energy}/E_F\equiv (e^2/\kappa r_0)/(\pi\hbar^2n/g_vm)\equiv r_s g_v$.
As stated before, $r_s\equiv r_e$ for $g_v=1$, but for $g_v>1$,
$r_e>r_s$, implying that the interpretation of the
quantum Coulomb coupling as a dimensionless energy would 
imply that increasing $g_v$ automatically involves a linear
increase of the dimensionless coupling parameter. In fact,
a third possible definition of the dimensionless coupling
strength is $r_x=\sqrt{g_v} r_s$, where $r_x$ now measures the
dimensionless strength of the average exchange energy to the average
kinetic energy at $T=0$. Again, $r_x$ increases with increasing
$g_v$ at a fixed $r_s$ implying that the quasiparticle
renormalization should be enhanced with $g_v$.
Although these alternative definitions of the Coulomb coupling
constant (e.g. $r_e$, $r_x$) have often been emphasized in the
literature \cite{abrahams} in order to claim that a multi-valley (i.e.
$g_v>1$) system is more strongly interacting than a single-valley
system at the same total electron density (i.e. fixed $r_s$), we
assert that the many-body problem for the multi-valley situation is
necessarily a two-parameter problem (i.e. $r_s$ and $g_v$) which can
not be described in any situation by a single effective parameter
such as $r_e\equiv g_v r_s$ or $r_x\equiv\sqrt{g_v}r_s$ or any other
combination of $r_s$ and $g_v$. The
non-trivial dependence of the interaction-induced many-body effects
on the two independent parameters $r_s$ and $g_v$, in particular,
how even the qualitative nature of the quasiparticle renormalization
as a function of $r_s$ and $g_v$ manifests itself completely
differently in different properties of the multivalley 2DEG (e.g
compressibility versus susceptibility), is the central theme of this
work. Specifically, we show that many-body effects could either be
enhanced or suppressed with increasing $g_v$.

Although the main motivation of our work is
theoretical, a part of our motivation comes from the extensive
recent experimental work on the multivalley AlAs 2D quantum wells
carried out at Princeton university \cite{Shayegan}. These experiments
demonstrate that the 
measured 2D spin susceptibility of AlAs quantum wells depends on
$g_v$ (as well as $r_s$), and in general the susceptibility is
smaller for larger values of $g_v$ with the difference between
$g_v=1$ and 2 decreasing with decreasing $r_s$ (i.e. increasing
density). It was emphasized in these experimental papers
\cite{Shayegan} that such a
higher value of susceptibility for $g_v=1$ compared with $g_v=2$ is
contrary to the popular wisdom, based on considerations which claim
$r_e(=g_v r_s)$ or $r_x(=\sqrt{g_v} r_s)$ to be the appropriate
interaction parameter, which would imply an
increasing many-body
renormalization  with increasing $g_v$. In the current
work, we resolve this puzzle by showing that the
many-body enhancement of the 2D spin susceptibility decreases with
increasing $g_v$ (between 1 and 2) except at very small values of
$r_s$. We also make the predication that an increasing $g_v$
will always increase the many-body renormalization effect of the
{\it compressibility} (for all values of $r_s$ and $g_v$) in a sharp
contrast with the susceptibility.

We employ the one-loop self-energy calculation as our basic
underlying theory to calculate the $T=0$ susceptibility and
compressibility as a function of $r_s$ and $g_v$. The self energy is
being calculated in the leading order expansion in the dynamically
screened Coulomb interaction because of the well-known long distance
(i.e. $q\rightarrow 0$) divergence of the bare Coulomb interaction.
This is equivalent to calculating the thermodynamic grand potential
in the infinite ring or bubble diagram expansion and then obtaining
the spin susceptibility and the compressibility by taking the
appropriate derivatives with respect to the magnetic field and the
volume respectively. The important thing to remember is that each
bubble diagram carries a factor of $g_v$ due to the valley
degeneracy (in addition to the factor of 2 due to the spin
degeneracy), and the Fermi wave vector $k_F$ and the Fermi energy
$E_F$ (i.e. the non-interacting chemical potential) are both
suppressed by the valley degeneracy: \be k_F=(2\pi n/g_v)^{1/2};
E_F=\pi\hbar^2n/(mg_v) \label{eq:1} \ee where $n$ is the 2D carrier
density and $m$ is the bare effective mass. The basic 2D
non-interacting  polarizibility function, i.e. the bubble diagram,
is given by, 
\be
\Pi(q,\omega)=-2g_v\int\frac{d^2k}{(2\pi)^2}
\frac{f_{k+q}-f_{k}}{\hbar \omega-[\varepsilon_{k+q}-\varepsilon_k]+i\delta} 
\label{eq:2} 
\ee 
where $\varepsilon_k=\hbar^2k^2/2m$, and $f_k$ is the Fermi
distribution function corresponding to the energy $E(k)$. We note
that the $q\rightarrow 0$, $\omega \rightarrow 0$ limit of the
bubble diagram gives the 2D non-interacting density of
states, $D(E) = g_v m /\pi \hbar^2$, which is enhanced by $g_v > 1$.
Eqs.~(\ref{eq:1}) and (\ref{eq:2}) show how $g_v$ enters the
theory: (1) By decreasing $E_F$ (Eq.~(\ref{eq:1})), $g_v$ tends to
enhance interaction effects; (2) by increasing screening through
Eq.(\ref{eq:2}), i.e. by enhancing the density of states,
$g_v$ tends to suppress the interaction effects.
Depending on which of these effects is more important in
determining a particular property, the many-body renormalization may
increase, decrease, or show a non-monotonic behavior with increasing
$g_v$.

We skip the technical details of the single-loop self-energy
calculation, and provide below the final formula 
we use for calculating the interacting susceptibility ($\chi$) and
compressibility ($K$) in terms of $r_s$ and $g_v$, 
\bea
\frac{\chi_0}{\chi}&=&1+\frac{\alpha r_s}{\sqrt{2}\pi}\int dq\int
du\frac{1}{\epsilon(q,iuq)} \nonumber\\
&&\times \left(\frac{d}{dk}
\frac{\sqrt{a+\sqrt{a^2+b^2}}}{\sqrt{a^2+b^2}}\right)
\Big{\vert}_{k=1}\label{eq:4}  \\
\frac{K_0}{K}&=&\frac{1}{k_F}
\frac{d(k_F^2/2+Re[\Sigma(k_F,E_F)])}{dk_F}\label{eq:5} 
\eea
In Eqs.~(\ref{eq:4}) and (\ref{eq:5}), $\chi_0$ and $K_0$ refer to
the corresponding non-interacting quantities. $a=u^2+k^2-q^2/4, b=uq$,
and $\epsilon({q,iuq})=1-v_q\Pi(q,iuq)$ is the dynamical RPA
dielectric
function, and $\Sigma(k_F,E_F)$ is the one-loop self energy calculated
at the Fermi surface, and
$\alpha=\sqrt{g_v/2}$ and $k_F=1/(\alpha r_s)$. 
We note that the final equations are
sufficiently complex that it is not possible to read off the
($r_s$, $g_v$) dependence for arbitrary values of $r_s$ and $g_v$.
Clearly, the results depend on both $r_s$ and $g_v$ independently.
We have, therefore,
numerically calculated the ($r_s$, $g_v$) dependence of the
susceptibility and the compressibility for arbitrary $r_s$ and $g_v$
values.

\begin{figure}[t]
\epsfxsize=.9\hsize
\hspace{0.0\hsize}
\epsffile{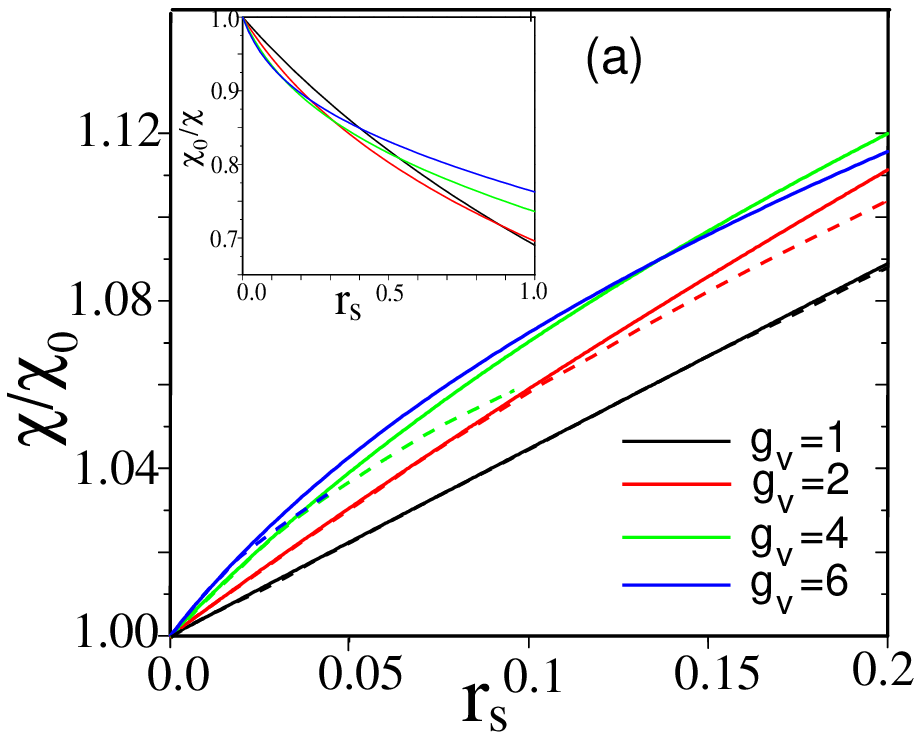}
\epsfxsize=.9\hsize
\hspace{0.0\hsize}
\epsffile{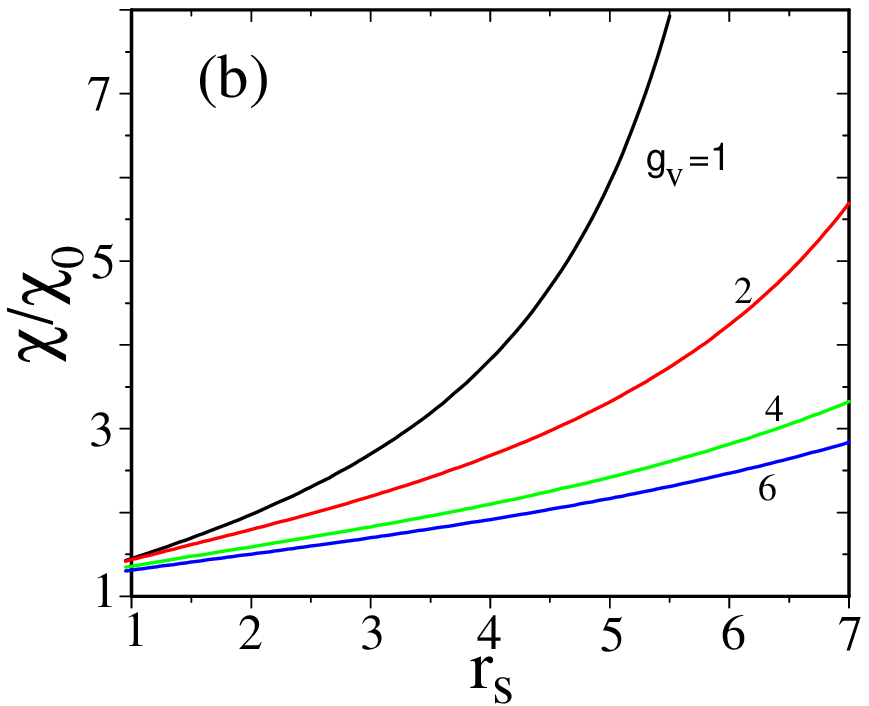}
\caption{(a) Susceptibility vs $r_s$ for different valley
degeneracy. The solid line is for the numerical data, dash-dotted
line for the asymptotic formula shown as Eq.~(\ref{eq:6}).
Inset gives
the inverse spin susceptibility vs $r_s \le 1$ showing
non-monotonicity.
(b) Susceptibility vs $r_s \ge 1$ for different valley degeneracy. 
} 
\label{fig:1a}
\end{figure}



\begin{figure}[t]
\epsfxsize=.9\hsize
\hspace{0.0\hsize}
\epsffile{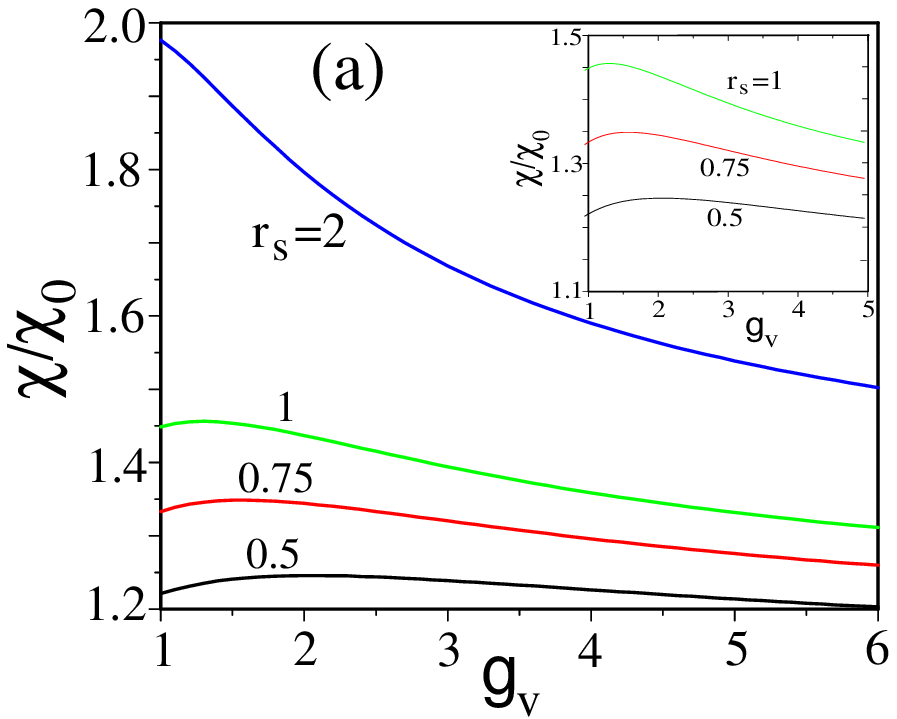}
\epsfxsize=.9\hsize
\hspace{0.0\hsize}
\epsffile{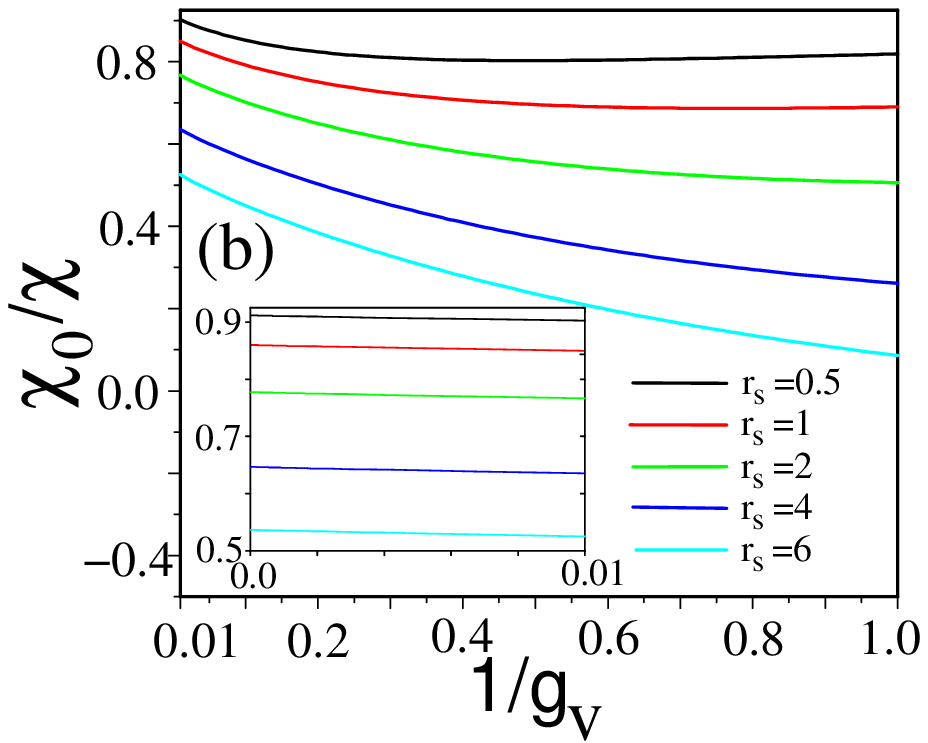}
\caption{(a) Susceptibility vs $g_v$ for different
electron density $r_s$. For small values, the
non-monotonicity can be seen in the inset. (b) Inverse susceptibility
vs $1/g_v$ for 
different $r_s$. 
Inset shows the constant $\chi$ for $g_v \gg 1$.}
\label{fig:2a}
\end{figure}



First, we present our susceptibility results in Fig.~1 and 2. In
Fig.~1, we show $\chi/\chi_{0}$ as a function of $r_s$ (for
$g_v=1,2,4,6$) in the small $r_s$ regime ($r_s=0 - 0.2$) in Fig.~1(a)
and in the large $r_s$ regime ($r_s=1-7$) in Fig.~1(b). It is clear
that the situation is qualitatively different between the small
($r_s<1$) and the large ($r_s>1$) $r_s$ regime. For small $r_s$, the
many-body effects are {\it enhanced} by increasing $g_v$ in
accordance with one's naive expectation based on the suppression of
$E_F$ by $g_v$. For large $r_s$ values, on the other hand, the
many-body renormalization of $\chi$ decreases with increasing $g_v$,
in complete agreement with the recent experimental measurements
\cite{Shayegan} which
were carried out in the $r_s>1$ regime. One of our predications is,
therefore, that a measurement of the 2D susceptibility at high
densities ($r_s<1$) would manifest a non-monotonic crossover in the
$g_v$ dependence of the many-body effects as can be seen in the inset
of Fig.1(a). In Fig.1(a), we show by the dotted line our analytic
formula valid in the $r_s \ll 1$ regime which can be derived to be
\be 
\frac{\chi}{\chi_0}=1+\sqrt{2g_v} \frac{r_s}{\pi} + \frac{g_v}{2}
\frac{r_s^2}{\pi^2} - 2.22 \frac{r_s^2g_v^2}{2\sqrt{2}\pi}.
\label{eq:6} 
\ee 
We can see from Eq.~(\ref{eq:6})
that $\chi$ can not be written, even in the $r_s\ll 1$ limit, as a
single-parameter function.

In Fig.~2, we show the $g_v$ dependence of the interacting
susceptibility, both for small and large $g_v$. The
non-monotonic behavior of the susceptibility as a function of $g_v$
for smaller $r_s$ values is obvious in Fig.~2, and the decrease
of $\chi$ with increasing $g_v$, eventually reaching a constant for
unphysically large $g_v$ (all $g_v>6$ is unphysical since no known
multivalley 2D semiconductor system has more than six valleys). For
$r_s < 1$, the maximum in $\chi/\chi_0$ occurs between
$g_v=1$ and 2, and should therefore be experimentally accessible
as a matter of principle. For $r_s>1$, however, the maximum in Fig.~2
moves to the $g_v<1$ regime which is unphysical. The very large
$g_v\gg6$ regime (Fig.~2b) is of theoretical interest only. We obtain
the following asymptotic analytic formula for $g_v\gg1$: 
\be
\frac{\chi}{\chi_0} \simeq 1+\frac{r_s^{2/3}}{4\times
2^{5/6}}+ \frac{1.4}{g_v} \simeq 1+0.14 r_s^{2/3}+ \frac{1.4}{g_v},
\label{eq:7} 
\ee 
up to logarithmic corrections in $r_s$ and $g_v$. Our
$g_v\rightarrow\infty$ formula (Eq.~(\ref{eq:7})) agrees well with
our numerical results for $g_v\gg100$. The small $g_v$
(and the small $r_s$) regime shown in Fig.~2 agrees with
the analytic Eq.~(\ref{eq:6}) derived for the small $r_s$ regime,
where the non-monotonicity with respect to $g_v$ can be seen to be
arising from the negative sign in the last term. We see that
Eq.~(\ref{eq:6}) implies a maximum in $\chi$ at $g_{v}^{m}\simeq1.07$
(for $r_s=1$) and 1.78 ($r_s=0.5$), which is consistent with the
numerical result of Fig.~2.

\begin{figure}[t]
\epsfxsize=.9\hsize
\hspace{0.0\hsize}
\epsffile{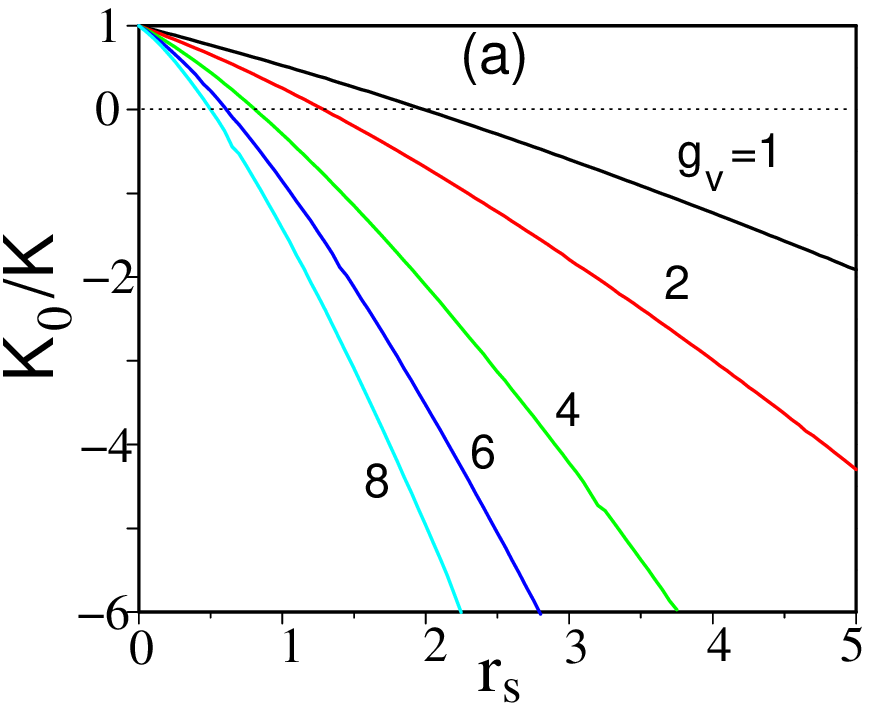}
\epsfxsize=0.9\hsize
\hspace{0.0\hsize}
\epsffile{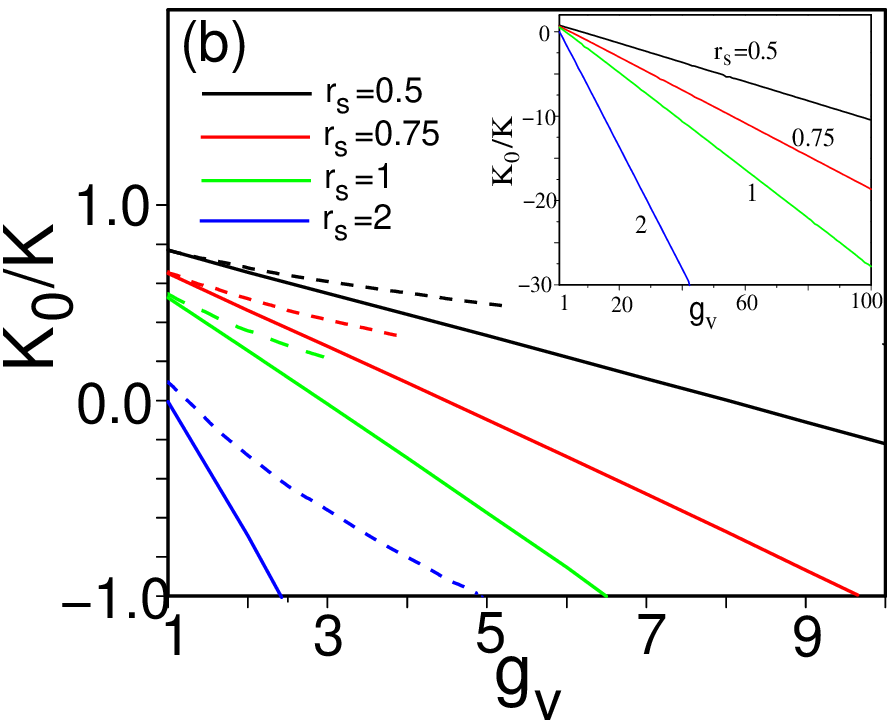}
\caption{(a)  Compressibility vs
$r_s$ for different valley degeneracy. 
(b) Compressibility vs $g_v$. Dotted lines are HF results. 
Inset shows the compressibility vs $g_v$ for large $g_v$.
%
}
\end{figure}


%

In Fig.~3, we show our calculated $r_s$ and $g_v$
dependence of the 2D interacting (inverse)
compressibility. It is
clear that, in contrast to the interacting susceptibility, the
interacting compressibility manifests monotonically stronger
many-body effects with increasing valley degeneracy $g_v$ in all
regimes of $r_s$ and $g_v$. There is, in fact, no
non-monotonicity at all in the interacting compressibility which
decrease continuously with increasing $r_s$ or $g_v$, eventually
becoming negative for $r_s\eqslantless2.5$ in a 2DEG
\cite{eisenstein}. We can 
analytically calculate the small $r_s$ dependence of the interacting
compressibility, obtaining: 
\be 
\frac{K_{0}}{K(r_s \ll 1)}\approx 1-0.45
r_s g_v^{1/2}+0.022(r_s^3g_v)\log(r_sg_v^{3/2}),
\label{eq:8} 
\ee
where we note that the first two terms of the expansion (i.e. the
Hartree-Fock result with just the exchange self-energy correction)
are identical to the corresponding first two terms (i.e. the
Hartree-Fock result) for $\chi_{0}/\chi$ in Eq.~(\ref{eq:6}). We
mention that the leading-order  correlation correction, the third
term in Eq.~(\ref{eq:8}), has an $r_s^3g_v$ dependence in the
compressibility (in contrast to the $r_s^2g_v^2$ dependence in
Eq.~(\ref{eq:6}) for the susceptibility).
Consequently the correlation correction to the compressibility is
very weak \cite{eisenstein} for small $r_s$ values. In Fig.~3(b), we
compare the 
Hartree-Fock result with our numerical results, and for small $r_s$,
the agreement is excellent. For large $g_v$, $K_0/K$ decreases
monotonically linearly in $g_v$ (neglecting logarithmic
corrections).

In discussing the significance of our results, we emphasize that we
have provided a complete qualitative resolution of the puzzle raised
by the experimental observations of ref. \cite{Shayegan}, where the
2D interacting susceptibility was found to be smaller for the larger
valley degeneracy system. In addition, we make a clear 
predication that if the experiments are
carried out in higher-density (i.e. lower $r_s$) samples than used
so far, the susceptibility will be larger in the higher valley
degeneracy system. We also predict that the many-body effects in the
2D compressibility, in sharp contrast to the 2D susceptibility,
would be enhanced monotonically with increasing $g_v$ for all $r_s$
values. We mention that in our theory the valley ($g_v$) and the spin
($g_s$) degeneracy are equivalent, and therefore a measurement of
valley susceptibility \cite{eight} in the presence of a variable spin
degeneracy (e.g. $g_s=2$ or 1) would manifest exactly the same
qualitative trend as seen in Figs.~ 1 and 2 with the roles of valley
and spin being interchanged. Similarly we predict a suppression of the
compressibility with decreasing spin degeneracy from $g_s=2$ to 1 by
applying an external parallel magnetic field at fixed value of $g_v$.
We believe that an
experimental confirmation of our qualitative predictions would be
compelling evidence of the Fermi liquid nature of 2D interacting
systems even at large $r_s$ and $g_v$ values.

This work is supported by DOE Sandia national Labs and by LPS-CMTC.


\begin{thebibliography}{99}

\bibitem{ando} T. Ando {\it et al}., \rmp {\bf 54}, 437 (1982).


\bibitem{mahan} G. D. Mahan, {\it Many Particle Physics}, (Plenum
  Publisher, New York, 2000); D. Pines and P. Nozieres, {\it Theory of
  Quantum Liquids}, (Solid 
State Physics, Academic Press, 1973).

\bibitem{lilly} M. P. Lilly {\it et al.}, 
Phys. Rev. Lett. {\bf 90}, 056806 (2003); J. Zhu {\it et al}., 
Phys. Rev. Lett. {\bf 90}, 056805 (2003);
M. J. Manfra {\it et al.}, 
Phys. Rev. Lett. {\bf 99}, 236402 (2007);
X. P. Gao {\it et al}., 
Phys. Rev. Lett. {\bf 93}, 256402 (2004);
Jongsoo Yoon {\it et al}., 
Phys. Rev. Lett. {\bf 82}, 1744 (1999).

\bibitem{ying} Ying Zhang and S. Das Sarma Phys. Rev. B {\bf 72},
075308 (2005); 
F. Perrot and M. W. Dharma-wardana, Phys. Rev. B {\bf 62}, 16536 (2000).
These two articles considered the many-body renormalization of 2D spin
susceptibility for $g_v=1$ and 2 theoretically, coming to conflicting
conclusions -- the first paper concluded that the renormalization is
larger for $g_v=1$ than for $g_v=2$ whereas the second one concluded
the opposite!


\bibitem{abrahams} E. Abrahams {\it et al}., 
Rev. Mod. Phys. {\bf 73}, 251 (2001).



\bibitem{Shayegan} 
Y. P. Shkolnikov {\it et al}., 
\prl {\bf 92}, 246804 (2004).
O. Gunawan {\it et al}., 
Phys. Rev. Lett. {\bf 97}, 186404 (2006).


\bibitem{eisenstein}
J. P. Eisenstein {\it et al}., 
Phys. Rev. Lett. {\bf 68}, 674 (1992); Phys. Rev. B {\bf 50}, 1760
 (1994); S. C. Dultz and H. W. Jiang,
Phys. Rev. Lett. {\bf 84}, 4689 (2000); S. Ilani {\it et al}., 
Phys. Rev. Lett. {\bf 84}, 3133 (2000); 
G. Allison {\it et al.}, 
\prl {\bf 96}, 216407 (2006). 


\bibitem{eight} M. Padmanabhan {\it et al}., \prb {\bf 78}, 161301
(2007). 


\end{thebibliography}
\end{document}